\documentclass{article}  

\usepackage{graphicx}
\usepackage{amsmath}
\usepackage{booktabs}
\usepackage{color}
\usepackage{authblk}

\begin{document}

\title{\textbf{Error correction in \\ quantum cryptography based on artificial neural networks}
}

\author{Marcin Niemiec       
}

\affil{AGH University of Science and Technology\\Mickiewicza 30, 30-059 Krakow, Poland\\\textit{niemiec@kt.agh.edu.pl}          
}

\date{ }

\maketitle

\begin{abstract}
Intensive work on quantum computing has increased interest in quantum cryptography in recent years. Although this technique is characterized by a very high level of security, there are still challenges that limit the widespread use of quantum key distribution. One of the most important problem remains secure and effective mechanisms for the key distillation process. This article presents a new idea for a key reconciliation method in quantum cryptography. This proposal assumes the use of mutual synchronization of artificial neural networks to correct errors occurring during transmission in the quantum channel. Users can build neural networks based on their own string of bits. The typical value of the quantum bit error rate does not exceed a few percent, therefore the strings are similar and also users' neural networks are very similar at the beginning of the learning process. It has been shown that the synchronization process in the new solution is much faster than in the analogous scenario used in neural cryptography. This feature significantly increases the level of security because a potential eavesdropper cannot effectively synchronize their own artificial neural networks in order to obtain information about the key. Therefore, the key reconciliation based on the new idea can be a secure and efficient solution.


\end{abstract}

\section{Introduction}
\label{intro}

Quantum cryptography is a technique which can ensure a very high level of data security. Thanks to principles of quantum mechanics, secret keys can be established between entities/users -- usually called Alice and Bob. At the same time, an eavesdropper (called Eve) can attempt to gain information about the key. However, measurement modifies the state of the transmitted information and even passive eavesdropping can be discovered by Alice and Bob.

After quantum key distribution in the quantum channel, the users must perform a key distillation process (consisting of quantum bit error estimation, error correction, and privacy amplification) in order to establish the secure final key. This process directly influences the performance of key distribution and also the security and length of final cryptographic key. Therefore, it is desirable to use secure and efficient methods in practical implementations. These features are inherent in the solution presented in this article -- synchronization of the artificial neural network to correct errors occurring in quantum channel during quantum key distribution process.

The rest of the article proceeds as follows. An introduction to quantum cryptography -- especially a description of the key distillation process -- is presented in Section 2. An introduction and presentation of the artificial neural networks used in neural cryptography follow in Section 3. The new solution based on machine learning in error correction process is presented in Section 4. In Section 5, experimental results are discussed and compared with typical scenarios used in neural cryptography. Finally, Section 6 concludes the article.

\section{Quantum cryptography}
\label{qc}

Quantum cryptography ensures secure key distribution by means of the laws of quantum mechanics \cite{ekert2000}.
First of all, the rules of quantum mechanics ensure that measurement modifies the state of the transmitted qubit (quantum bit). This modification can be discovered by the sender and the receiver of qubits. Therefore, quantum cryptography requires two types of channels to be defined:
\begin{itemize}
\item[$\bullet$]\textbf{the quantum channel,}
where qubits with the information about the distributed key are exchanged and

\item[$\bullet$]\textbf{the public channel,}
which is used to check whether the communication through the quantum channel is distorted. Also, this channel is used for the correction of wrong bits.
\end{itemize}

The other rule of quantum mechanics which makes quantum cryptography a very secure solution is the no-cloning theorem~\cite{zurek82}.
According to this theorem, it is not possible to create identical copies of an unknown quantum state~\cite{lo08}. Therefore, an eavesdropper is not able to clone the original qubit to measure the quantum state and send the second qubit to the proper receiver.

We can split quantum cryptography into two main steps: the quantum key distribution protocol (e.g. BB84 protocol) and the key distillation algorithms (quantum error bit estimation, error correction and privacy amplification). 

\subsection{Quantum key distribution}
\label{qkd}

Quantum key distribution (QKD) is used to distribute an encryption key for symmetric ciphers~\cite{miralem2017} (not to transmit messages between users). As has been mentioned, the security of QKD relies on the foundations of quantum mechanics and information about a key is transmitted by means of qubits. We could distinguish two types of QKD protocols: based on single and entangled particles~\cite{dusek06}.

In the first group -- QKD protocols based on single particles -- information about the distributed key is coded by means of quantum states of single particles (such as polarized photons). The quantum states of the particles do not depend on each other and each particle brings information which can be read independently.

The second group is based on entanglement. The entangled state of two particles has the following feature: the states of particles are random (indeterminate) before the measurement is performed but if we measure the state of the first particle, then the state of the second particle is fully determined. This means that we only need to measure one particle to know the states of them both. It is worth mentioning that the entanglement still retains this feature even if the particles are separated.

Today, we know a lot of QKD protocols but only a few are used in practice~\cite{scarani09}. The first protocol invented was BB84~\cite{bb84}, presented in 1984 by Bennett and Brassard. This protocol is based on single particles (polarized photons).
Another protocol based on single particles is B92 -- developed by one of the creators of BB84, Bennett, in 1992~\cite{b92}. It is simpler and faster than its predecessor. Furthermore, it is more efficient because it detects eavesdroppers faster.
A well-known QKD protocol based on entanglement is E91, invented in 1991 by Ekert \cite{ekert91}. It was an innovative solution which used the phenomenon of entangled particles for the first time.
In principle, many other protocols, such as BBM92 \cite{bbm92} (proposed by Bennett, Brassard and Mermin in 1992) or SARG04 \cite{sarg04} (proposed by Scarano, Acín, Ribordy, and Gisin in 2004) are modified versions of the BB84 protocol.

\subsection{Key distillation}
\label{kd}

During the quantum key distribution process, Alice and Bob use two communication channels: quantum and public. In the quantum channel, information is coded by means of quantum states. In the public channel Alice and Bob exchange data to check whether Eve is eavesdropping. However, the public channel is necessary for more cases.

It is not only Eve that is responsible for errors in the quantum channel. Errors during quantum communication may occur because of disturbance in the quantum channel, optical misalignment, noise in detectors, or other factors. Therefore, Alice and Bob have to estimate the error rate and decide if there is an eavesdropper in between or not.
In practice, they compare a small portion of a distributed raw key through the public channel and compute the quantum bit error rate (QBER). The portion of compared bits can depend on the security requirements \cite{niemiec13}.
If QBER exceeds a given threshold, it means that Eve has eavesdropped (or the quantum channel is too noisy to perform a proper key distribution). But if the error rate is low enough, Alice and Bob continue further distillation of the key. Of course, they must delete the compared part of the raw key for security reasons.

After the bit error estimation, Alice and Bob use key distillation protocols. These protocols usually involve two steps: key reconciliation (error correction) and privacy amplification.

As mentioned previously, quantum communication is not perfect and some errors usually occur. If the number of errors does not exceed a given threshold of QBER, the reconciliation process must find and correct or delete these errors. Alice and Bob should disclose as little information as possible by using an appropriate reconciliation algorithm. Since they are not able to avoid the leakage of information, they have to reject some bits of the key. 

The first binary error correction method was provided by the BBBSS protocol. This protocol was designed by Bennett and his coworkers~\cite{bennett92}. It requires the parities of raw key subsets from Alice and Bob to be exchanged. BBBSS uses several passes to correct the errors by parity check. A pseudo-random permutation is used after each pass.
Two years later, Brassard and Salvail constructed the Cascade algorithm with improved efficiency \cite{brassard94}. Usually, it uses four passes and doubles block length starting from the second pass. This ensures a faster error correction process. Nowadays, the Cascade key reconciliation algorithm is usually used in practical implementations.
Other reconciliation methods based on the BBBSS algorithm are Furukawa-Yamazaki \cite{furukawa01} (less efficient than the Cascade) and Winnow protocol \cite{buttler03} which uses a Hamming code to reduce the number of errors.

Alice and Bob can choose one of several known reconciliation algorithms; however, currently the most popular reconciliation methods are algorithms which are based on a parity check of blocks. The simplest scenario assumes that the key is grouped into blocks of a given size. The size of a block depends on the error rate value which was estimated before. Alice and Bob compare
parities of each block over the public channel. If their parities disagree, the block contains an odd number
of errors. This block is cut into two sub-blocks and their parities are compared again. This procedure is continued recursively for all blocks
which contain an odd number of errors as long as errors will be corrected. After that, both keys contain an even number of errors or
none. Alice and Bob shuffle the positions of bits and repeat the same procedure with blocks of bigger size as long as both keys will be the same.
A serious problem occurs if blocks contain an even numbers of errors. Therefore, users must try to change  the block size or rearrange the position of errors in the string. However, this approach can be ineffective and may even lead to failing the error correction process.

Each parity control over the public channel discloses a part of the secret key's information. If Eve collects the parities of many blocks, she will be able to calculate parts of the key. Therefore, Alice and Bob must reject some bits to reduce the eavesdropper's knowledge about the secret key. Many
rejected bits increase the security level but unfortunately decrease the final length of the key. It decreases the efficiency of whole QKD system. The ideal key reconciliation algorithm should ensure an efficient and secure error correction process as well as avoid leakage of information about the key.

At the end of the key distillation process, the privacy amplification should be carried out. Because Eve may have gained significant knowledge of the key (eavesdropping in the quantum channel and in the public channel during the bit error estimation and key reconciliation), Alice and Bob are required to strengthen their privacy. They can delete some of the bits and construct the final key in a specific way. 

Even though it is possible to apply different solutions during the privacy amplification process, universal hash functions are mainly used in practice. Universal families of hash functions were created by Wegman and Carter \cite{carter79}. Privacy amplification with hash functions was proposed by Bennett, Brassard and Robert \cite{bennett88} in 1988. In general, the algorithm is based on one-way functions which are able to convert a large string of bits into a short binary word.

Following \cite{bennett88}, the theorem which defines the probability of the eavesdropper's information after error correction is presented below.
\\
\\
\textbf{Theorem 1.}
\emph{Assume that $M$ is the length of the reconciled key and Eve's knowledge about the key is no more than $E$ deterministic bits.
Let $h : \{0,1\}^{M} \rightarrow \{0,1\}^{E}$ be any hash function of the universal family, let $S<M-E$ be a security parameter,
and let $R = M-E-S$. If $g : \{0,1\}^{M} \rightarrow \{0,1\}^{R}$ is chosen randomly, then the
expected amount of information on $g(x)$ given by knowledge of $h$, $g$ and $h(x)$ is at most: $2^{-S} / \log2$.
This means that:}
\begin{equation}\label{s}
    \text{expected amount of information}\,\, \leq \,\,\frac{2^{-S}}{\log2} \,\, [bit].
\end{equation}

The security parameter $S$ allows the security of the final encryption key to be controlled.
By means of the theorem, we are able to increase the security of a given QC system.
Unfortunately, too many rejected bits decreases the final length of the key, and thus the efficiency of QKD system also decreases.

If Alice and Bob perform all these steps, the final key will be significantly reduced. This
is characteristic for all quantum key distribution protocols \cite{mehic15}. 
Because each stage reduces the key length, the performance of QKD is also reduced. Sometimes, when we want to ensure a high level of security, this reduction is significant. Using the QKD Protocol Simulator \cite{symulatorQKD} we can easily check that e.g. 1000 qubits transmitted in the quantum channel cause approx. 300 bits of final key. Therefore improving efficiency of key distillation process is crucial to the quantum cryptography implemented in real communication networks.

\section{Artificial neural  networks}

Artificial neural  networks (ANN)  are  a  family  of  statistical  learning  models  inspired  by  biological  neural  networks \cite{ann2}. They are used to estimate functions that can depend on a large  number  of  inputs.
An ANN consists of artificial neurons (analogous to biological neurons) which are connected together. Each connection can transmit a signal between neurons \cite{ann1}.
Neurons are usually organized in layers: the first layer consist of input neurons which can send the data to the second layer (called hidden). A neural network can have one or more hidden layers. The last layer -- consisting of output  neurons -- is called the output layer. 
The connections can store parameters (called weights) that can be manipulated during calculation. 

\subsection{Tree Parity Machine}
\label{tpm}
\begin{figure*}
  \includegraphics[width=1\textwidth]{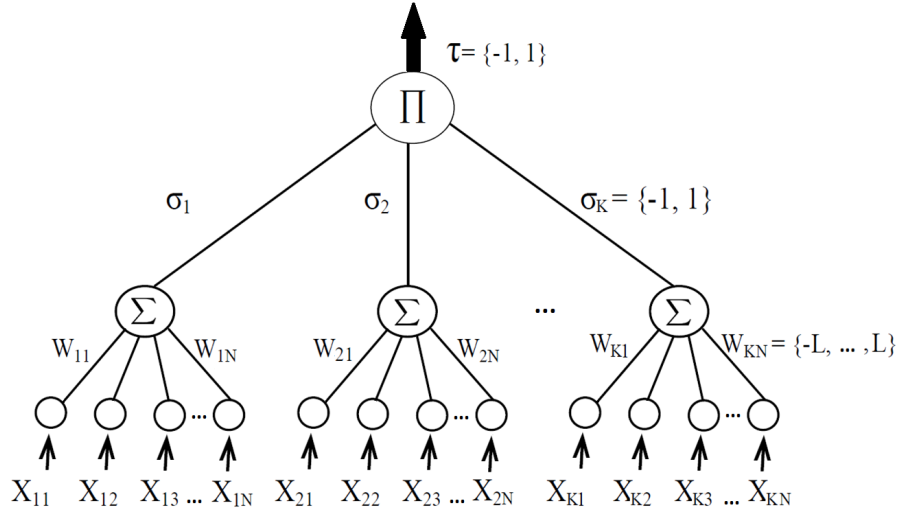}
\caption{TPM machine}
\label{fig:1}       
\end{figure*}

The most popular neural network used for cryptography purposes is the tree parity machine (TPM) which contains only one hidden layer. An example TPM structure is presented in Figure \ref{fig:1}. 
It consists of $K N$ input neurons, where $K$ is the number of neurons in the hidden layer and $N$ is the number of inputs into each neuron in the hidden layer. This network has only one output neuron.
Each connection between the input layer and hidden layer is characterized by its weight, which is an integer from the range $ [-L, L] $.

The output value of neuron $k$ in the hidden layer depends on input $x$ and weight $w$ and is calculated as:
\begin{equation}
\sigma_{k}=sgn(\sum_{n=1}^{N} x_{kn}\ast w_{kn})
\end{equation}
where signum function is:
\begin{equation}
sgn(z)=  \begin{cases}
    -1       & z\leq0\\
    1        & z>0
  \end{cases}
\end{equation}
The output value of the neuron in the output layer is calculated as:
\begin{equation}
\tau=\prod_{k=1}^{K} \sigma_{k}
\end{equation}

Nowadays, the TPMs are used for establishing the secret key between users. This usage of ANN for cryptographic purposes is called neural cryptography. Alice and Bob use two identical neural networks which are able to synchronize after mutual learning \cite{ann3}. At the beginning of this process, each TPM generate random values of weights but after synchronization process both users have TPMs with the same values of weights. Therefore, Alice and Bob can construct the secret key using synchronized weights (just change weight values into binary string).

In order to synchronize neural networks, users generate random input (the same for both TPMs) and compute outputs from each TPM. If the output of Alice's TPM is the same as Bob's TPM, they can start the learning process for the neural networks. If the outputs are different (one TPM generated the value $1$ but the other generated the value $-1$), Alice and Bob must generate another input.

We can choose any learning algorithm; however, the generalized form of Hebbian method is the most popular in practical implementations \cite{hebb}. The new weights are calculated by means of the following formula:
\begin{equation}
 w_{kn}^{\star} = \nu_{L}(w_{kn} + x_{kn} \ast \sigma_{k} \ast \Theta(\sigma_{k}, \tau ))
\end{equation}
where:
\begin{equation}
\Theta(\sigma_{k}, \tau ))=  \begin{cases}
    0        & \quad \text{if }  \sigma_{k} \neq \tau \\
    1        & \quad \text{if }  \sigma_{k} = \tau
  \end{cases}
\end{equation}
and function $ \nu_{L} $ limits values of connections to the range $[-L, L]$:
\begin{equation}
\nu_{L} (z) =  \begin{cases}
    -L        & \quad \text{if }  z \leq -L \\
    z        & \quad \text{if }   -L < z < L \\
    L        & \quad \text{if }   z \geq L 
  \end{cases}
\end{equation}
As we can see, the algorithm strengthens the connections which have the same value as the TPM output.

After the appropriate number of iterations, the synchronization process ends and the weights of both TPM machines are the same. Then, Alice and Bob can change weights into binary strings and use them as a secret cryptographic key.

\subsection{Security of neural cryptography}
\label{learn}

\begin{figure*}
  \includegraphics[width=1\textwidth]{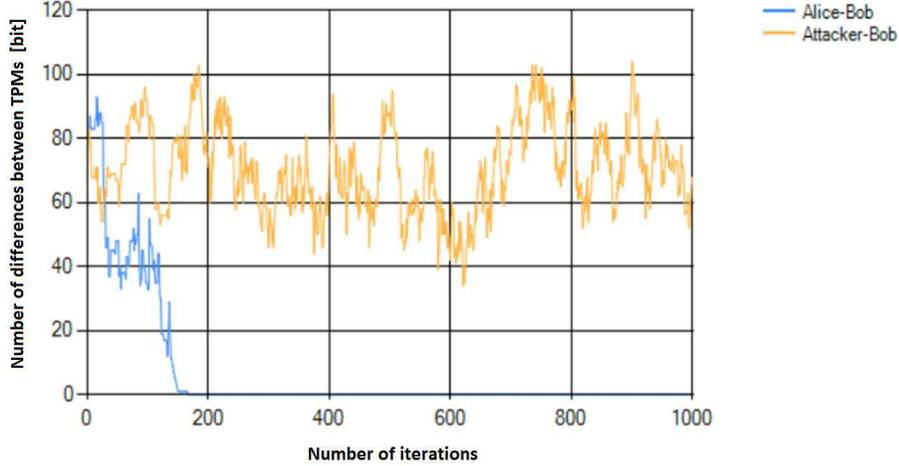}
\caption{Synchronization of TPMs (Alice, Bob and intruder)}
\label{fig:2}       
\end{figure*}

Synchronization of TPMs requires communication between Alice and Bob. Therefore, it can be eavesdropped by an intruder (Eve). The simplest passive attack is an attempt to synchronize the Eve's TPM machine with the TPMs belonging to Alice and Bob. We can specify that during the synchronization process, three events may occur:
\begin{itemize} 
\item[*] if $\tau^{Alice} \neq \tau^{Bob}$, then no TPM machine is subjected to the learning process,
\item[*] if $\tau^{Alice} = \tau^{Bob} \neq \tau^{Eve}$, then only the machines of Alice and Bob are subjected to the learning process,
\item[*] if $\tau^{Alice} = \tau^{Bob} = \tau^{Eve}$, then all machines are subjected to the learning process.
\end{itemize} 
If the output of the Eve's TPM machine is different than the outputs of the Alice and Bob's machines, the learning process cannot be performed. Therefore, the synchronization of the Eve's TPM is slower than the synchronization of the TPMs belonging to Alice and Bob.
An example of the synchronization process is presented in Figure \ref{fig:2} (TPM machines with parameters: N=8, K=6, L=2 and Hebbian learning algorithm). Alice and Bob synchronized neural networks before $200$ iterations, but attacker was not able to do it for $1000$ iterations.

Unfortunately, an attack can be improved by using multiple TPM's owned by Eve. In this case, the attacker has many machines, each initiated with different weights. This method increases the probability of the attacker's success, since it is enough that only one machine will be synchronized with Alice's and Bob's machines. However, simultaneous synchronization of many TPM machines is ineffective and users can easily improve the security by increasing the used neural network (increasing L, N and K parameters).
This results in a reduction in the synchronization speed of the attacker.

It is worth mentioning a known improvement to the introduced simple passive attack. The event $\tau^{Alice} = \tau^{Bob} \neq \tau^{Eve}$ cannot stop Eve's synchronization process. This approach is known as a geometric attack~\cite{sec-1}. The attacker can flip the output of a selected neuron in the hidden layer before applying the learning process in order to correct the output $\tau^{Eve}$. Taking into account the correlation of weights in TPMs machines belonging to Alice, Bob and Eve, the authors of~\cite{sec-2} considered attractive and repulsive steps of the synchronization process. Despite the fact that a geometric attack can improve the learning process of Eve's TPM machine, the synchronization process is still less effective. Finally, it was demonstrated that ANNs which interact with each other (Alice's and Bob's TPMs) synchronize faster than Eve's TPM machine.

The security of neural cryptography has been considered in a number of papers~\cite{ann2,sec-11,sec-22,r3-1,r3-2,r3-3,r3-4}. However, synchronization of TPMs can be further improved by learning by queries~\cite{sec-3} instead of random inputs. This approach is based on exchanging inputs between Alice and Bob which are correlated to the weight vectors of the TPMs. The queries introduce a mutual influence between Alice and Bob which is not available to an attacking Eve. The results shown in~\cite{sec-4} confirm that queries restore the security against cooperating attackers.

\section{Error correction based on TPMs}

One of the crucial step in the quantum key distribution process is the correction of errors. This step decides on the security level of final key, but also significantly influences the performance of the quantum cryptography. Currently used solutions assume parity checking and deleting bits to minimize the probability of information leakage. Such an approach causes a significant reduction of key length and low efficiency in the whole QKD system. In this Section, a new approach to error correction based on mutual synchronization of TPM machines is introduced.

\subsection{Applicability of TPMs}
\label{app-of-TPMs}

The idea for a new error correction method is the following: after the QBER estimation step, we can use the synchronization of the TPM machines to correct errors in the quantum cryptography (instead of any other error correction algorithm). In this way, we will be able to correct errors that occurred during the transmission of qubits. Importantly, in this scenario, Alice's binary string is very similar to Bob's string of bits. The typical value for QBER does not exceed a few percent~\cite{qkd-example-1,qkd-example-2,qkd-example-3,qkd-example-4}, therefore we must correct only a small part of the whole key. This means that the TPM machines are close to synchronization and the learning process will finish much faster than in the case of synchronization of random strings of bits. Of course, this increases the security level significantly.

It is worth mentioning that the presented idea -- using the mutual synchronization of neural networks to correct errors -- is a special case when this process makes sense. In general, TPM machines cannot be used for error correction of digital information because we are not able to predict the final weights after the learning process. Both TPM machines dynamically adjust their weights, therefore the final strings are random. However, in the case of quantum cryptography this feature is an advantage because we want to generate unpredictable string of bits which can be used as a secure key.

Taking into account the software environments and hardware resources currently available, the usage of TPM-based error correction seems to be justified. Although the time of learning processes for software implementations of TPMs strongly depends on the hardware resources, a typical duration of one iteration on an ordinary computer is a few milliseconds~\cite{ann4}. However, hardware implementation helps to shorten this time by more than thousandfold (i.e. parallel processing using FPGA presented in~\cite{impl-1}). 
Additionally, ASIC implementations of neural cryptography in 130nm and 65nm standard-cell CMOS technologies are available~\cite{impl-2}. These circuits reduce implementation costs and ensure fast synchronization of neural networks; the maximum operation frequency is several hundred MHz which results less than one microsecond per single iteration.

Additionally, the security services and architectures being used confirm the feasibility of TPM implementation in practice. For example, a chip-level microcomputer bus systems with TPMs introduced in~\cite{impl-3} provides efficient data encryption with a low hardware overhead, comparable to well-known and widely used stream ciphers. Moreover, synchronization of ANNs was proposed to environments such as ad-hoc networks (TPMs for establishing common group keys~\cite{impl-4}) or wireless sensor networks with limited resources (lightweight key agreement protocol based on TPMs known as TinyTPM~\cite{impl-5}).

\subsection{Error correction process}
\label{error-corr-proc}

The use-case with the proposed solution is as follows.
Let's assume that Alice and Bob carried out the process of quantum key distribution in the quantum channel and they estimated the quantum bit error rate. If the QBER level is acceptable (this means that no one eavesdropped on the quantum channel or a very small percentage of bits were eavesdropped), the error correction process can start.

\begin{description}
\item[\textbf{Step 1}] \hfill \\
Alice and Bob create their own TPM machines based on their own strings of bits. The users change string of bits into weights in their own TPM machines (bits into numbers from the range $ [-L, L] $ ). Values $ \{ -L, -L+1, ... L-1, L \} $ become weights of connections between the input neurons and the neurons in the hidden layer. Values of parameter $K$ (the number of neurons in the hidden layer) and $N$ (the number of inputs into each neuron in the hidden layer) are chosen by Alice and Bob and can be public.
In this way, Alice and Bob construct very similar neural networks -- the TPM machines have the same structure ($K N$ input neurons) and most of the weights are the same. The differences are located only in the places where errors occurred: for example, if QBER $ \approx 3\% $, it means that $ \approx 97\% $ of bits are correct.
\\
\item[\textbf{Step 2}] \hfill \\
After the construction of the neural networks, synchronization of the TPM machines begins and continues until all weights in both machines become the same. 
In order to synchronize neural networks, Alice generates random input (an input string has $KN$ length) and computes output from her own TPM machine. Then, Alice informs Bob about the result (value $1$ or $-1$) and also the generated input string. Bob computes output from his own TPM machine, based on the input string generated by Alice. If the output of Alice's TPM is the same as Bob's TPM, they can start the learning process for the neural networks (otherwise, the different input string is generated by Alice). 
The synchronization process can be based on the Hebbian learning algorithm, which strengthens the connections which have the same value as the TPM output.
After the appropriate number of iterations, the synchronization process ends and the weights of both TPM machines are the same.
\\
\item[\textbf{Step 3}] \hfill \\
When the TPM machines are synchronized, the weights are the same in both neural networks. Therefore Alice and Bob can convert the weights back into string of bits -- the users change numbers from the range $ [-L, L] $ into bits (in the opposite way than in Step 1).
Because the both TPM machines have been synchronized, the Alice's string of bit is now the same as the Bob's string of bits. All errors have been corrected.
\\
\end{description}

After this three-step error correction process, both users may use the obtained string of bits for cryptography purposes, e.g. to secure communications ensuring confidentiality, integrity or authentication.

\subsection{Security considerations}
\label{security-consider}

The presented solution -- using the mutual synchronization of TPM machines to correct errors -- is not based on parity check which causes the information leakage, however, privacy amplification process is still recommended. It will protect this solution against unknown attacks to TPMs, which can be proposed in the future.

Let's assume that before the error correction process Alice and Bob changed string of bits into weights and created their own TPM machines. Thus, the TPM machines contains $K N$ input neurons with weights and each weight is an integer from the range $ [-L, L] $. Therefore, a single weight has $2 L + 1$ possible values. The number of possible keys, which are stored using TPM is:
\begin{equation}
 (2 L + 1) ^ {K N}
\label{eq-8} 
\end{equation}
However, after each synchronization process (iteration) Eve is able to acquire a partial information about TPMs. Taking into account the input and output of TPMs (value $\tau$), Eve can reject the half of possible keys (equation~\ref{eq-8}) form the further considerations. Therefore, after the $i$ iterations, the number of possibilities is reduced to:
\begin{equation}
 2 ^ {- i} (2 L + 1) ^ {K N}
\label{eq-9} 
\end{equation}
and this is adequate to a TPM machine with smaller number of input neurons with weights:
\begin{equation}
 (2 L + 1) ^ {K N - Z}
\label{eq-10} 
\end{equation}
Comparing both equations (\ref{eq-9} and \ref{eq-10}) we are able to quantify the maximum Eve's knowledge after $i$ iterations and define the reduction of key to protect Alice and Bob against the information leakage during TPMs synchronization process as:
\begin{equation}
 Z = log _ {(2 L + 1)} 2 ^ i
\label{eq-11} 
\end{equation}
This reduction strongly depends on the parameter $L$. However, the reduction of key caused by the synchronization process for the typical QBER value is not very high. Even for small parameter $L$ -- i.e. TPMs with $L = 2$ used for verification in the next section) -- the reduction is a dozen or so percent.
When Alice and Bob convert the weights back into bits, they may shorten the final key using a hash function and a proper value of the security parameter $S$ (regarding Equation \ref{s}). In this way, Alice and Bob reduce Eve's knowledge of the key, which can be collected by eavesdropping in the quantum channel and public channel during the bit error estimation and key reconciliation steps.

Additionally, the proposed solution is characterized by higher security than current neural cryptography solutions, where we use TPM machines to establish cryptography key between users. In the new solution we have much faster synchronization because the strings are very similar at the beginning of the synchronization process. Therefore, Alice and Bob need fewer iterations to synchronize their neural networks. 
However, error correction based on TPMs is able to equalize every number of incompatible bits between Alice and Bob's strings of bits and it works on any value of QBER.
Also, it is worth remembering that we should use high values for TPM parameters (N, K, L) to ensure an even higher level of protection \cite{ann4}.

\section{Verification}

Security and efficiency of every key reconciliation method are the crucial requirements. It also applies to the proposed error correction based on mutual synchronization of artificial neural networks.
The number of iterations during the synchronization of TPMs influences the security and efficiency. If the synchronization of Alice and Bob's TPMs is fast, the level of security will be high. Therefore, the scenarios of TPMs synchronization process with typical values of QBER are tested in this Section.

\subsection{The security of error correction based on TPMs}

The number of steps during the synchronization of TPMs directly influences the security level. However, the synchronization of an eavesdropper's TPM machine is slower than the synchronization of the users' TPMs, but the initial synchronization additionally increases the level of security. To verify the behaviour of the initial synchronized TPMs, a number of simulations have been conducted. The results were compared to typical TPMs used in neural cryptography (artificial neural networks with random chosen weights). 

\begin{figure*}
  \includegraphics[width=1\textwidth]{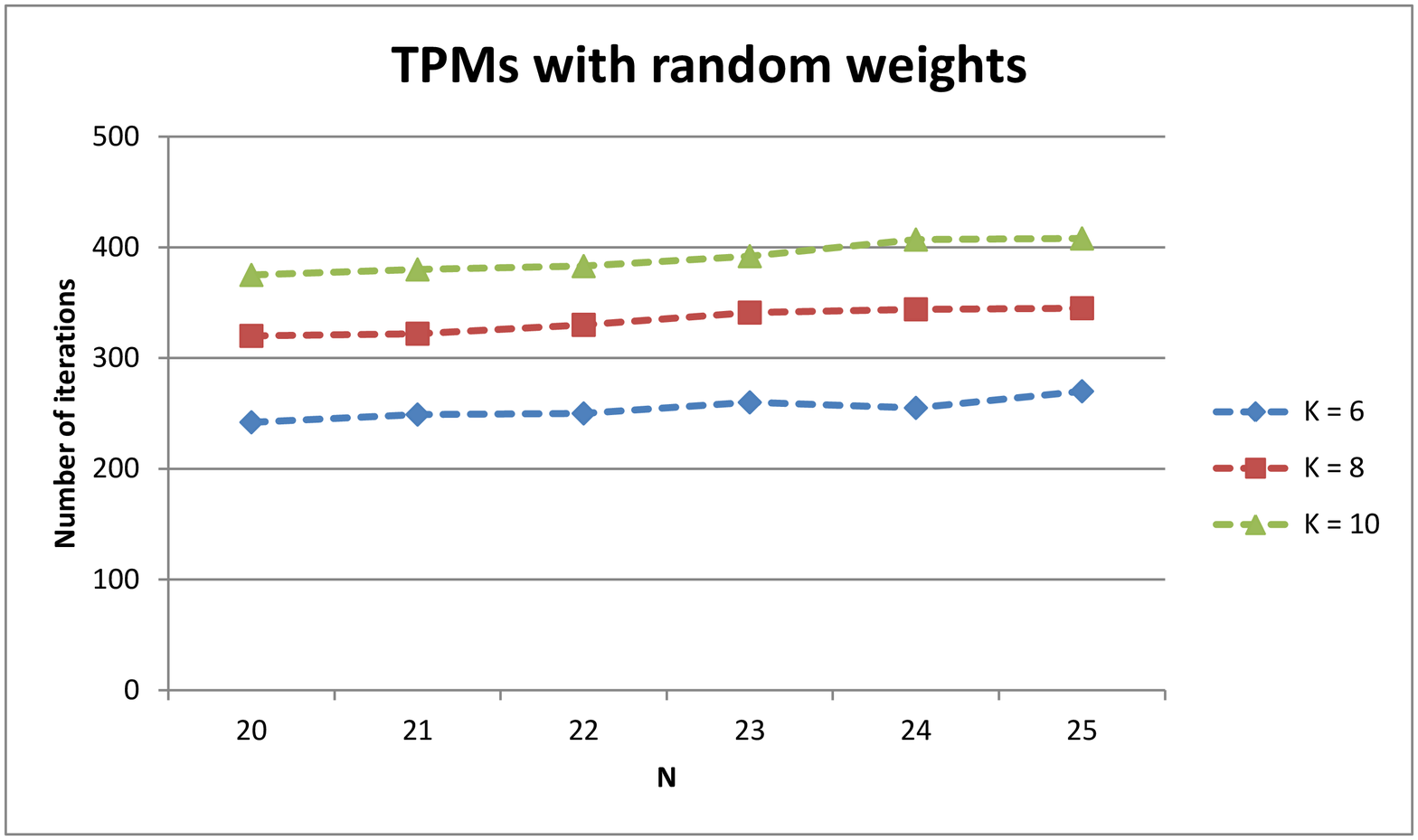}
\caption{Synchronization of TPMs with random chosen weights}
\label{fig:3}       
\end{figure*}

\begin{figure*}
  \includegraphics[width=1\textwidth]{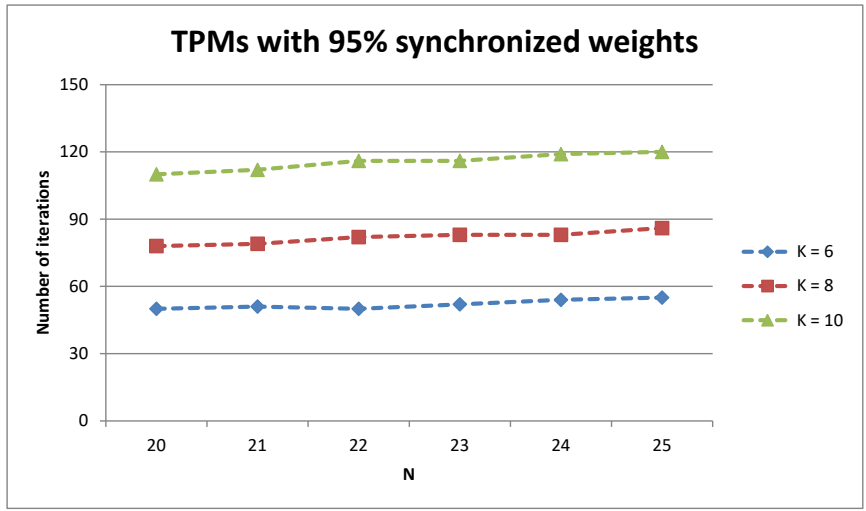}
\caption{Synchronization of TPMs with 95\% synchronized weights at the beginning of the synchronization process}
\label{fig:4}     
\end{figure*}

Figures \ref{fig:3} and \ref{fig:4} present the synchronizations of TPMs in two scenarios -- with weights randomly generated and with 95\% synchronized weights at the beginning of the synchronization process, respectively. The points in the graph are mean values (synchronizations repeated $5000-10000$ times) and were connected by dotted lines in order to help in comparison of differences. The results were presented for the range $N=[20,25]$, parameter $L=2$ and Hebbian learning algorithm.
The number of iterations in synchronization process strongly depends on the value of the $K$ parameter (the figures contain three example values: $K=6$, $K=8$ and $K=10$). 

According to predictions, the number of iterations which are needed to synchronize TPMs is much smaller in scenario with ANN initially synchronized ($3-4$ times smaller). Artificial neural networks with random chosen weights need significantly more iterations to synchronize their weights. 

\begin{figure*}
  \includegraphics[width=1\textwidth]{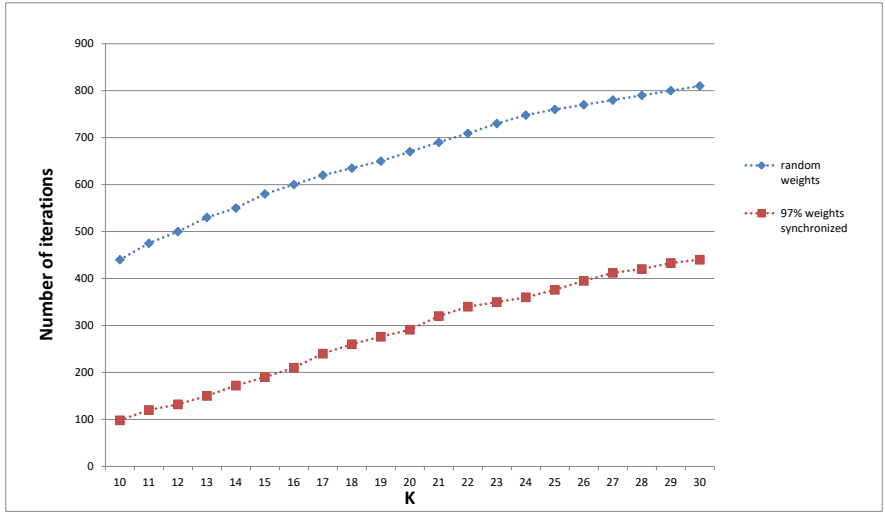}
\caption{Synchronization of TPMs with parameter $N = 30$ ($L=2$, Hebbian learning algorithm)}
\label{fig:5}      
\end{figure*}

\begin{figure*}
  \includegraphics[width=1\textwidth]{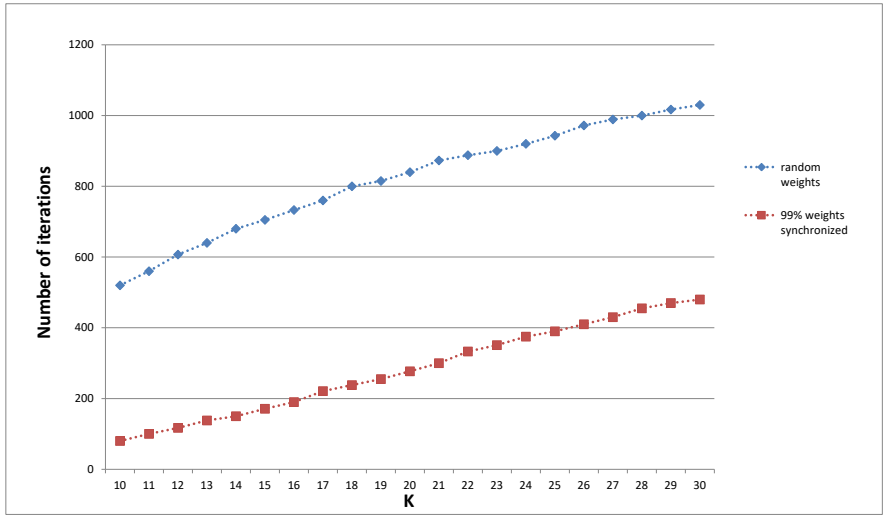}
\caption{Synchronization of TPMs with parameter $N = 50$ ($L=2$, Hebbian learning algorithm)}
\label{fig:6}      
\end{figure*}

Additionally, numerous simulations were performed with synchronization of bigger ANNs in scenarios with QBER = 3\% and QBER = 1\%. Figure \ref{fig:5} presents the comparison of speed of TPMs synchronization depending on parameter $K$ in two scenarios: with random weights and with 97\% synchronized weights at the beginning of the TPMs synchronization process (both for $N = 30$). Figure \ref{fig:6} presents results for bigger TPMs ($N = 50$) and allows differences between the synchronization of TPMs with random chosen weights and TPMs with very similar weights (differences at 1\%) to be compared.

All the presented scenarios confirm that initially synchronized TPMs require far fewer iterations than artificial neural networks with random chosen weights. This feature increases the level of security significantly. The proposed solution is much more secure than neural cryptography where TPMs are used to establish cryptographic keys using random strings at the beginning of synchronization process.

\subsection{The efficiency of error correction based on TPMs}

In order to compare the efficiency of the new TPM-based approach with other error correction algorithms (BBBSS~\cite{bennett92} and Cascade~\cite{brassard94}), additional tests were performed. Following recommendation~\cite{brassard94}, it was assumed that the block size for the BBBSS algorithm should have $\frac{0.73}{QBER}$ bits. The blocks in the first pass of the Cascade algorithm are of the same length. Tests were performed for different distribution of errors in the key. It was also assumed that both algorithms corrected all errors after four passes.
Verifying the efficiency of the tested algorithms, two different keys were chosen: 
\begin{itemize}
\item[$\bullet$] \textit{key length} $= 500$ bits with $QBER = 5\%$ and 
\item[$\bullet$] \textit{key length} $= 600$ bits with $QBER = 3\%$. 
\end{itemize}
The results are presented in Table~\ref{table1}. The table contains the average number of iterations for each tested error correction algorithm. In the case of BBBSS and Cascade algorithms, the iteration means a parity check of a single block. The values for TPM-based error correction come from Figure \ref{fig:4} (TPM parameters: $N=25$, $K=10$, $L=2$) and Figure \ref{fig:5} (TPM parameters: $N=30$, $K=10$, $L=2$). For both key lengths tested, the number of iterations is significantly lower for the new TPM-based approach than the BBBSS and Cascade algorithms.

\begin{table}
\caption{Comparison of the error correction algorithms}
\label{table1}
\centering
\begin{tabular}{ | p{3.5cm} || p{2cm} | p{2cm} | p{2cm} |}
\toprule
 & \begin{center} \textbf{BBBSS} \end{center} & \begin{center} \textbf{Cascade} \end{center} & \begin{center} \textbf{TPM-based} \end{center} \\
\midrule
\midrule
\begin{center} \textbf{Number of iterations} \\ \textit{(Key lenght = 500 bits} \\ \textit{QBER = 5\%)}  \end{center} & \begin{center} 213 \end{center} &	\begin{center} 181 \end{center} & \begin{center} 120 \end{center} \\
\hline
\begin{center} \textbf{Number of iterations} \\ \textit{(Key lenght = 600 bits} \\ \textit{QBER = 3\%)}  \end{center} & \begin{center} 189 \end{center} & \begin{center} 150 \end{center} & \begin{center} 98 \end{center} \\
\bottomrule
\end{tabular}
\end{table}

\section{Conclusions}

In this article, a new idea for the key reconciliation method in quantum cryptography is presented. The proposal assumes that artificial neural networks (TPM machines) can be used to correct errors occurring in the quantum channel. Unlike neural cryptography, the new solution is characterized by fast synchronization of TPM machines. Typical values of QBER do not exceed a few percent, therefore users need to correct only a small part of the key. This means that the TPM machines used for this purpose are close to synchronization and the learning process can finish quickly. 

It was shown that the synchronization process in the new solution is much faster than in the case where TPMs weights are chosen randomly (the typical scenario of neural cryptography). 
When the QBER value does not exceed a few percent, the synchronization process is several times faster.
This significantly increases the level of security because of the problem with fast synchronization of eavesdropper's TPMs which must start from randomly generated weights. Therefore, the key reconciliation based on the synchronization of TPM machines can be a secure and efficient solution. 
The presented solution can replace the error correction algorithms currently used in the quantum key distribution process. 

Although the risk of information leakage in the proposed solution is low, additional mechanisms should be used to increase the level of security. First of all, the use of the privacy amplification process after the error correction process is still recommended. Also, users can consider dividing a long key into shorter strings and perform error correction processes separately for each string. After that, the privacy amplification process should be performed on the whole key (concatenated by all the strings). This approach decreases the risk of information leakage, even if an eavesdropper could get some information about a selected string.

Although TPM machines are not used for error correction of digital information, this approach can be used for key reconciliation in quantum cryptography. Mutual synchronization of TPMs dynamically adjust their weights, therefore the final weights are not predictable. Fortunately, this is a big advantage of key distribution, because a secure key for cryptographic purposes should be a random string of bits. It is a very special case when artificial neural networks can be used to correct errors.

Additionally, it is worth mentioning that error correction based on TPMs is resistant to currently known attacks using a quantum computer. This feature is likely to be particularly important in the near future.
\\
\\
\\
\textbf{Acknowledgements}
This work was funded by the Polish National Centre for Research and Development under "SDNRoute: integrated system supporting routing in Software Defined Networks" project  no. LIDER/30/0006/L-7/15/NCBR/2016.

\end{document}